# BAYESIAN COMPARISON OF GARCH PROCESSES WITH SKEWNES MECHANISM IN CONDITIONAL DISTRIBUTIONS *

Mateusz Pipień

Department of Econometrics, Cracow University of Economics

The main goal of this paper is an application of Bayesian model comparison, based on the posterior probabilities and posterior odds ratios, in testing the explanatory power of a set of competing GARCH (ang. Generalized Autoregressive Conditionally Heteroscedastic) specifications, all with asymmetric and heavy tailed conditional distributions. In building competing volatility models we consider, as an initial specification, conditionally Student-t GARCH process with unknown degrees of freedom parameter, proposed in [7]. By introducing skewness into Student-t family and incorporating the resulting class as a conditional distribution we generated various GARCH models, which compete in explaining possible asymmetry of both conditional and unconditional distribution of financial data. In order to make Student-t family skewed we consider various alternative mechanisms recently proposed in the literature. In particular, we apply the hidden truncation mechanism (see [3], [1]), an approach based on the inverse scale factors in the positive and the negative orthant (see [10]), order statistics concept ([14]), Beta distribution transformation ([15]) and Bernstein density transformation (see [28]). Additionally, we consider GARCH process with conditional $\alpha$-Stable distribution, see [30], [29]. Based on the daily returns of hypothetical financial time series, we discuss the results of Bayesian comparison of alternative skewing mechanisms applied in the initial Student-t GARCH framework. Additionally, we present formal Bayesian inference about conditional asymmetry of the distribution of the daily returns in all competing specifications on the basis of the skewness measure defined in [2].

* Presented at 2nd Symposium on Socio- and Econophysics, Cracow 21−22 April 2006 Research supported by a grant from Cracow University of Economics. The author would like to thank Malgorzata Snarska for help in manuscript preparation in Latex format

# 1. Introduction

The presence of both, conditional and unconditional skewness (asymmetry) of the distributions of the financial time series returns has been recognized for decades. But, as suggest [19], only a few attempts to specify formally this feature have been made. A proper modeling of skewness in the distribution of financial returns is important at least for two reasons. Firstly, uncaptured skewness clearly affects inference about all parameters of the sampling model. As a consequence the final conclusions, drawn from the sampling model which does not allow for asymmetry can be misleading. Lanne and Saikkonnen present in [19] the impact of the conditional skewness assumption on the results of making inference about the volatility and expected return. They presented empirical analysis, which showed, that a positive and significant relation between return and risk can be uncovered, once an appropriate probability distribution is employed to allow for conditional asymmetry. Motivating the importance of asset pricing model that incorporates conditional asymmetry, [16] emphasize that systematic skewness is economically important and commands a risk premium. Investigating the influence of the assumption of asymmetric distributions in portfolio selection, [18] concluded, that, if investors prefer right-skewed portfolios, then for equal variance one should expect a "skew premium" to reward investors willing to invest in left-skewed portfolios. Secondly, in pricing the derivatives and in risk management, the accurate models, which describe the return process are particularly desired. The importance of the assumption of conditional skewness in models used for option pricing was presented [31], [17] and [8]. Additionally, conditional skewness clearly influences the results of risk assessment built on the basis of the Value at Risk (VaR) concept. Application of time varying volatility models with conditional asymmetric distributions in Value at Risk prediction present [9]; for a Bayesian approach to VaR calculation see [29].
Within GARCH (Generalised Autoregressive Conditionally Heteroscedastic) framework, initially proposed by [6] as a conditionally normal stochastic process, fat tailed and possibly asymmetric distributions have been also proposed and applied. Osiewalski and Pipień in [26] defined GARCH process with conditional skewed Student-t distribution, which is an asymmetric generalisation of Student-t family proposed by [10]. In [21], [20] and [30] GARCH process with conditional $\alpha$-Stable distribution was considered. Some other processes with asymmetric conditional distribution were applied in [16], [30], [32], and [9]. Despite of the fact, that many researchers found the conditionally skewed volatility models better than those, which do not allow for asymmetry, there is very hard to find the result of the formal comparison of explanatory power of such specifications. Many authors con-clude the superiority of conditional skewed models on the basis either of the asymptotically based statistical significance of the skewness excess (see e.g. [32], or [19]) or of informal likelihood inference (see e.g. [30], [32]). Hence more formal approach to investigating the explanatory power of conditionally skewed models seems to be necessary. Additionally, the results of formal comparison of competing unnested specifications of conditional skewness could be very valuable in selection the best skewing mechanism.

On the other hand, in recent years in statistics it can be noticed a peculiar interest in the theory and applications of distributions that can account for skewness. This resurgent field of research yields new families of possibly asymmetric sampling models, as well as more general methods of measuring skewness phenomenon. The most common approach to the creation of the family of skewed distributions is to introduce skewness into an originally symmetric family of distributions. This approach underlies the general classes of skewed probability distributions generated for example by hidden truncation mechanism (see [3], [1]), inverse scale factors applied to the positive and the negative orthant (see [10]), order statistics concept ([14]), Beta distribution transformation ([15]), Bernstein density transformation (see [28]) and the constructive method recently proposed by [11].

The main goal of this paper is to define a set of competing GARCH specifications, all with asymmetric conditional distributions, which also allow for heavy tails. As an initial specification we consider GARCH model with conditional Student-t distribution with unknown degrees of freedom parameter, proposed by [7]. By introducing skewness, according to the methods mentioned above, and by incorporating the resulting family as a conditional distribution, we generate GARCH models which compete in explaining possible asymmetry of the conditional and unconditional distribution of the financial data. We also consider GARCH process with conditional $\alpha$-Stable distribution, which, from the definition, also allows for skewness, see [25].

By application of Bayesian approach to model comparison, based on the posterior probabilities and posteriori odds ratios, we test formally the explanatory power of competing, conditionally fat tailed and asymmetric GARCH processes. Based on the daily returns of hypothetical financial time series, we discuss the results of Bayesian comparison of alternative skewing mechanisms and also check the sensitivity of model ranking with respect to the changes in prior distribution of model specific parameters. Additionally we present formal Bayesian inference about conditional asymmetry in all competing specifications on the basis of the skewness measure defined in [2].

## 2. Creating asymmetric distributions

Let consider parametric family of absolute continuous real random variables $I = \{\varepsilon_f; \varepsilon_f : \Omega \to \mathbb{R}\}$, parameterized by the vector $\theta$. For each value of $\theta \in \Theta$, by $f(.|\theta)$ and $F(.|\theta)$ we denote the density and cumulative distribution function (cdf) of $\varepsilon_f$. Let assume, that for each $\theta \in \Theta$ the density $f(.|\theta)$ is unimodal and symmetric around the mode. Consider another parametric family $P$ of absolute continuous random variables, which distributions are defined over the unit interval, $P = \{\varepsilon_f; \varepsilon_f : \Omega \to (0,1)\}$, with density $p(.|\eta_p)$ parameterized by vector $\eta_p \in H$. The unified representation of univariate skewed distributions that we study in this paper is based on the inverse probability transformation. In our approach the class $I$ is the initial family of symmetric distributions, while the class $P$ defines formally skewing mechanism. The family of absolute continuous random variables $IP = \{\varepsilon_s, \varepsilon_s : \Omega \to \mathbb{R}\}$, with general form of density $s(.|\theta, \eta_p)$ is said to be the skewed version of the symmetric family $I$, if the density $s$ is given by the form:

$$s(x|\theta, \eta_p) = f(x|\theta) \cdot p\left(F(x|\theta)|\eta_p\right), \quad for \quad x \in \mathbb{R} \qquad (1)$$

A number of simple but very powerful results can be obtained from decomposition (1); see [11]. The most important and rather intuitive fact is that the distributions $s$ and $f$ are identical if and only if $p(.|\eta_p)$ is the density of the uniform distribution over the unit interval; i.e. if $p(y|\eta_p) = 1$, for each $y \in (0,1)$. Hence if we want to create the family of distributions $IP$ such that $I \subset IP$, we must assure, that the uniform distribution over $(0,1)$ can be obtained in family $P$ for some specific value $\eta_p^* \in H$.

Within the general form (1) several classes of distributions $P$ have been considered and incorporated into some specific families of symmetric random variables in order to obtain skewness. The first approach of making distribution $F(.|\theta)$ skewed applied hidden truncation ideas. The skew-Normal distribution in [3] constitutes the first explicit formulation of such a mechanism. In general this approach assumes, that:

$$s(x|\theta, \gamma_2) = 2 \cdot f(x|\theta) F(\gamma_2 \cdot x|\theta), \quad for \quad x \in \mathbb{R} \qquad (2)$$

where $\gamma_2 \in \mathbb{R}$ is the only one parameter which governs the skewing mechanism; $\eta_p = (\gamma_2)$. In this case, it can be shown, that $p(y|\gamma_2) = 2F(\gamma_2 F^{-1}(y)|\theta)$, for $y \in (0,1)$. In (2) positive and negative values of $\gamma_2$ define right and left skewed distributions. Since, for each $y \in (0,1)$, it is true that $p(y|0) = 2F(0F^{-1}(y)|\gamma_2) = 1$, the case $\gamma_2 = 0$ retrieves symmetry. As an alternative it was proposed in [14] to apply the family of Beta distributions in order to define $p(.|\eta_p)$. In particular, $s(x|\theta, \gamma_3)$ can be defined as follows:

$$s(x|\theta, \gamma_3) = f(x|\theta) Be\left(F(x|\theta)|\gamma_3, \gamma_3^{-1}\right), \quad for \quad x \in \mathbb{R} \qquad (3)$$

where $Be(y|a, b)$ is the value of the density function of the Beta distribution with parameters $a > 0$ and $b > 0$, calculated at $y \in (0, 1)$. Since $Be(.|1, 1)$ defines the density of the uniform distribution, we obtain, that for $\gamma_3 = \gamma_3^{-1} = 1$ the density $s$ is symmetric. In (3) there is still only one parameter $\gamma_3 > 0$, which defines the type of asymmetry. If $\gamma_3 > 1$, then $s$ is right asymmetric, while $\gamma_3 < 1$ constitutes left asymmetric density.

The family $IP$ of skewed distributions proposed in (3) can be generalized, by imposing Beta distribution transformation with two free parameters $a > 0$ and $b > 0$. This leads to the following form for $s$

$$s(x|\theta, \eta_p) = f(x|\theta)Be(F(x|\theta)|a, b), \quad for \quad x \in \mathbb{R} \qquad (4)$$

In this case the vector $\eta_p = (a, b)$ contains two parameters, which govern skewness. As a consequence such a mechanism enables to vary tail weight. If $a = b = 1$ we go back to symmetry, while $a < b$ or $a > b$ defines left or right skewness. It can be shown that the skewing mechanism (4), in case when $I$ is the family of Student-$t$ distributions, yields skewed Student-$t$ family of distributions proposed in advance in [15].

Another method for introducing skewness into an unimodal distribution is based on the inverse scale factors on the left and on the right side of the mode of the density $f(.|\theta)$. Investigating this concept Fernandez and Steel proposed in [10] skewed Student-t family of distributions with the density $f_{sks}(.|\nu, 0, 1, \gamma_1)$ defined as follows:

$$f_{sks}(x|\nu, 0, 1, \gamma_1) = \frac{2}{\gamma_1 + \gamma_1^{-1}}\{f_t(x\gamma_1|\nu, 1, 0)I_{(-\infty,0)} + f_t(x\gamma_1^{-1}|\nu, 1, 0)I_{(0,+\infty)}\}$$

where $f_t(z|\nu, 1, 0)$ is the value of the density function of the Student-$t$ distribution with $\nu$ degrees of freedom, zero mode and unit inverse precision, calculated at $z \in \mathbb{R}$. The approach studied in [10] can be applied to any family $I$ of symmetric distributions by defining in (1) the following skewing mechanism for each $y \in (0, 1)$:

$$p(y|\gamma_1) = \frac{2}{\gamma_1 + \gamma_1^{-1}} \frac{\{f(\gamma_1 F^{-1}(y))I_{(0;0.5)} + f(\gamma_1^{-1} F^{-1}(y))I_{(0.5;1)}\}}{f(F^{-1}(y))}, \qquad (5)$$

where $\gamma_1 > 0$. The resulting density $s(.|\theta, \gamma_1)$ is symmetric if $\gamma_1 = 1$, while $\gamma_1 > 1$ or $\gamma_1 < 1$ make distribution right or left skewed.

As pointed in [11] the general form of density $s$ in (1) seems to be the good starting point in completely nonparametric treatment of the skewing mechanism $p$. As $\varepsilon_p : \Omega \to (0, 1)$ can be in general any random variable with probability distribution defined over the unit interval, the possibility to model it in an unrestricted fashion is tempting. The next approach of constructing $p$ is a compromise between totally flexible skewing mechanism and one obtained in parametric fashion. It uses Bernstein densities (see e.g. [28]), which are convex discrete mixtures of appropriate densities of Beta distributions. The following form on $p$ constitutes another skewing mechanism:

$$p(y|w_1,\ldots,w_m) = \sum_{j=1}^{m} w_j Be(y|j, m-j+1) \, , \, y \in (0,1)$$

where $m > 0$, $w_j \geq 0$, $w_1 + \ldots + w_m = 1$. The resulting $s(.|\theta, \eta_p)$ takes the form:

$$s(x|\theta, \eta_p) = f(x|\theta) \cdot \sum_{j=1}^{m} w_j Be(F(x|\theta)|j, m-j+1) \ \ for\ x \in \mathbb{R} \qquad (6)$$

where $\eta_p = (w_1,\ldots,w_{m-1})$, $w_j \in (0,1)$ for $j = 1,\ldots, m-1$, and in (6) $w_m = 1 - w_1 - \ldots - w_{m-1}$. For any $m > 0$, if $w_j = m^{-1}$, for each $j = 1,\ldots, m-1$, then Bernstein density reduces to the uniform distribution case. Hence equal weights $w_j$ lead to the symmetry in (6).

In the next section we present basic model framework, which is a starting point in generating conditionally heteroscedastic models for daily returns. In order to create the set of competing specifications, we make use of all presented skewing mechanisms. We also consider GARCH process with conditional $\alpha$-Stable distribution.

### 3. Basic model framework and competing skewed conditional distributions

Let denote by $x_j$ the value of a currency at time $j$. Following [4],[5], [27] let consider an AR(2) process for $\ln x_j$ with asymmetric GARCH(1,1) error. In terms of logarithmic growth rates $y_j = 100\ln(x_j/x_{j-1})$ our basic model framework is defined by the following equation:

$$y_j - \delta = \rho(y_{j-1} - \delta) + \delta_1 \ln x_{j-1} + \varepsilon_j \qquad j = 1, 2, \ldots \qquad (7)$$

The AR(2) formulation adopted from [5] enables to make inference on the presence of a unit root in $\ln x_j$. If $\delta_1 = 0$, then (7) reduces to the AR(1) process for $y_j$, i.e. an $I(1)$ process for $\ln x_j$. In an initial specification $M_0$ we assume, that the error term $\varepsilon_j = z_j(h_j)^{0.5}$, where $z_j$ are independent, Student-$t$ random variables, with $\nu > 0$ degrees of freedom parameter, mode $\zeta_1 \in (-\infty, +\infty)$, and unit inverse precision; i.e. $z_j \sim iiSt(\nu, \zeta_1, 1)$. The density of the distribution of the random variable $z_j$ is given as follows:

$$p(z|M_0) = f_t(z|0, 1, \nu) = \frac{\Gamma(0.5(\nu+1))}{\Gamma(0.5\nu)\sqrt{\pi\nu}} \left[1 + \frac{(z-\zeta_1)^2}{\nu}\right]^{-(\nu+1)/2} \qquad (8)$$

Defining $h_j$ we follow GJR-GARCH$(1,1)$ specification proposed in [12]:

$$h_j = a_0 + a_1 \varepsilon_{j-1}^2 I(\varepsilon_{j-1} < 0) + a_1^+ \varepsilon_{j-1}^2 I(\varepsilon_{j-1} \geq 0) + b_1 h_{j-1} \quad j = 1, 2, \ldots \quad (9)$$

which allows to model asymmetric reaction of conditional dispersion measure $h_j$ to positive and negative sign of shock $\varepsilon_{j-1}$.
As a consequence, in model $M_0$, the conditional distribution of $\varepsilon_j$ (with respect to the whole past of the process, $\psi_{j-1} = (\ldots, \varepsilon_{j-2}, \varepsilon_{j-1})$) is a Student-$t$ distribution with $\nu > 0$ degrees of freedom parameter, mode $\zeta_1 \in (-\infty, +\infty)$, and inverse precision $h_j$; i.e. $\varepsilon_j | \psi_{j-1}, M_0 \sim iiSt(\nu, \zeta_1, h_j)$. In specification $M_0$ the conditional distribution of $y_j$ is the Student-$t$ distribution with $\nu > 0$ degrees of freedom parameter, mode $\mu_j = \delta + \rho(y_{j-1} - \delta) + \delta_1 \ln x_{j-1} + \zeta_1 h_j^{0.5}$ and inverse precision $h_j$ (given by the equation (9)):

$$p(y_j | \psi_{j-1}, M_0, \theta, \nu) = f_t(y_j | \mu_j, h_j, \nu), \quad j = 1, 2, \ldots$$

where $\theta = (\delta, \rho, \delta_1, a_0, a_1, a_1^+, b_1, h_0)$ is the vector of all parameters defined in sampling model $M_0$ except the degrees of freedom parameter $\nu$.
Now we want to construct a set of competing GARCH specifications $\{M_i, i = 1, \ldots, k\}$ by introducing skewness into conditional distribution of $y_j$ in $M_0$. The resulting asymmetric distributions are obtained by skewing the distribution of the random variable $z_j$, (8), according to methods presented in the previous section. The resulting skewed density of $z_j$ is of the general form given by (1):

$$p(z | M_i) = f_t(z | 0, 1, \nu) p[F_t(z - \zeta_1) | \eta_i, M_i], \quad for \; z_j \in \mathbb{R}, \; i = 1, 2, \ldots, k$$

where $p(.|\eta_i, M_i)$ defines the skewing mechanism parameterized by the vector $\eta_i$, and $F_t(.)$ is the cumulative distribution function of the Student-$t$ random variable with $\nu > 0$ degrees of freedom parameter, zero mode and unit inverse precision. The resulting conditional distribution of $\varepsilon_j$ in model $M_i$ takes the form:

$$p(\varepsilon_j | \psi_{j-1}, M_i) = f_t(h_j^{-0.5}(\varepsilon_j - \zeta_1) | 0, 1, \nu) h_j^{-0.5} p[F_t(h_j^{-0.5}(\varepsilon_j - \zeta_1)) | \eta_i, M_i]$$

where $f_t(.|0, 1, \nu)$ is defined by the formula (8). This leads to the general form of the conditional distribution of daily return $y_j$ in model $M_i$:

$$p(y_j | \psi_{j-1}, \theta, \nu, \eta_i, M_i) = f_t(z_j^* | \nu, 0, 1) h_j^{-0.5} p[F_t(z_j^* | \eta_i, M_i], \quad (10)$$

where $z_j^* = h_j^{-0.5}(\varepsilon_j - \mu_j)$. As the first specification, namely $M_1$, we consider GARCH model with skewed Student-$t$ distribution obtained by the method proposed in [10]. The skewing mechanism $p[.|\eta_1, M_1]$ is given by the formula (5), where $\eta_1 = \gamma_1 > 0$, and $\gamma_1 = 1$ defines symmetry (i.e. $M_1$ reduces to the model $M_0$ under restriction $\gamma_1 = 1$). The model $M_2$ is the result of skewing conditional distribution $p(y_j|\psi_{j-1}, M_0, \theta)$ according to the hidden truncation method. In this case $p[.|\eta_2, M_2]$ is defined by (2), $\eta_2 = \gamma_2 \in \mathbb{R}$, while $\gamma_2 = 0$ defines symmetric Student-$t$ conditional distribution for $y_j$. In model $M_3$ we apply [14] Beta skewing mechanism with one asymmetry parameter. The skewing distribution $p[.|\eta_3, M_3]$ is defined by (3), where $\eta_3 = \gamma_3 > 0$, and $\gamma_3 = 1$ reduces our model to the case of $M_0$. Specification $M_4$ is based on the Skewed Student-$t$ distribution proposed by [15]. In this case $p[.|\eta_4, M_4]$ is defined by the formula (4), $\eta_4 = (a,b)$, for $a > 0$ and $b > 0$ and $a = b = 1$ reduces $M_4$ to $M_0$. In model $M_5$ we apply Bernstein density based skewing mechanism with $m = 2$ free parameters. It means that the skewing mechanism $p[.|\eta_5, M_5]$ is defined by the formula (6) and $\eta_5 = (w_1, w_2)$. The case $w_1 = w_2 = 1/3$ defines symmetry of the conditional distribution of $y_j$, given $M_5$.

As an alternative for all methods of making family of Student-$t$ random variables skewed, it is possible to consider in a GARCH framework a class of distributions, which directly, from the definition, enables for fat tails and skewness. The next GARCH specification is based on the assumption of conditional $\alpha$-stability. In GARCH model $M_6$, as a specification which is not a direct generalization of model $M_0$, we considered in (7) conditional $\alpha$-Stable distribution. In particular we put $\varepsilon_j = z_j(h_j)^{0.5}$, where $z_j$ are independent $\alpha$-Stable random variables with $\alpha \in (0,2]$, location parameter $\zeta_1 \in (-\infty, +\infty)$, unit scale and skewness parameter $\beta \in [-1,1]$; i.e. $z_j \sim iiSta(\zeta_1, 1, \beta, \alpha)$. For a report of Bayesian inference in model $M_6$ see [29].

We denote by $y^{(t)} = (y_1, \ldots, y_t)$ the vector of observed up to day $t$ (used in estimation in day $t$) daily growth rates and by $y_f^{(t)} = (y_{t+1}, \ldots, y_{t+n})$ the vector of forecasted observables at time $t$. The following density represents the $i$-th sampling model ($i = 1, 2, 3, 4, 5, 6$) at time $t$:

$$p(y^{(t)}, y_f^{(t)}|\theta, \omega_i, \eta_i, M_i) = \prod_{j=1}^{t+n} p(y_j|\psi_{j-1}, \theta, \omega_i, \eta_i, M_i) \quad i = 1, \ldots, 6,$$

where $\omega_i$ is the vector of additional parameters of the sampling model, which are not included in $\theta$ and $\eta_i$; for each $i = 1, 2, 3, 4, 5$ $\omega_i = \nu$, while $\omega_6 = \alpha$. The sampling model $M_i$ is based on the product of the appropriate densities $p(y_j|\psi_{j-1}, \theta_i, \omega_i, M_i)$, which are generally specified in the formula (10) for $i = 1, 2, 3, 4, 5$, while in case $i = 6$ $p(y_j|\psi_{j-1}, \theta, \omega_6, \eta_6, M_6)$ is defined by the appropriate density of $\alpha$-Stable distribution (see [29]).

Constructed at time $t$ Bayesian model $M_i$, i.e. the joint distribution of the observables $(y^{(t)}, y_f^{(t)})$ and the vector of parameters $(\theta, \omega_i, \eta_i)$ takes the form:

$$p(y^{(t)}, y_f^{(t)}, \theta, \omega_i, \eta_i|M_i) = p(y^{(t)}, y_f^{(t)}|\theta, \omega_i, \eta_i, M_i) p(\theta, \omega_i, \eta_i|M_i) \quad (11)$$

and requires formulation of the prior distribution $p(\theta, \omega_i, \eta_i | M_i)$, for each specification $M_i$, for $i = 1, 2, 3, 4, 5, 6$. In general we assumed the following prior independence:

$$p(\theta, \omega_i, \eta_i | M_i) = p(\theta | M_i) p(\omega_i | M_i) p(\eta_i | M_i) \quad i = 1, 2, \ldots, 6 \qquad (12)$$

The prior information about the common parameters $\theta$ was initially formulated by [27]. For $i = 1, 2, 3, 4, 5$ the prior density $p(\omega_i | M_i) = p(\nu | M_i)$ defines exponential distribution with mean 10 for the degrees of freedom parameter $\nu$. In case of conditionally $\alpha$-Stable GARCH model ($i = 6$) the density $p(\omega_6 | M_6) = p(\alpha | M_6)$ defines the uniform prior distribution over the interval $(0, 2]$ for the index of stability $\alpha$. For $i = 1$, $\eta_1 = \gamma_1 > 0$, and $p(\eta_1 | M_1)$ is the density of the standardized lognormal distribution truncated to the interval $\gamma_1 \in (0.5; 2)$. For $i = 2$, $\eta_2 = \gamma_2 \in \mathbb{R}$, and $p(\eta_2 | M_2)$ is the density of the normal distribution with zero mean and variance equal to 3. For $i = 3$, $\eta_3 = \gamma_3 > 0$, and $p(\eta_3 | M_3)$ is the density of the standardized lognormal distribution. In case of $i = 4$, $\eta_4 = (a, b)$, and $p(\eta_4 | M_4)$ is the product of the densities of the standardized lognormal distribution. For $i = 5$, $\eta_5 = (w_1, w_2)$ and $p(\eta_5 | M_5)$ is the product of the normal densities, both with mean 0.33 and variance 36, truncated by the following set of restrictions: $w_1 > 0, w_2 > 0, w_1 + w_2 < 1$. For $i = 6$ $\eta_6 = \beta$, and $p(\eta_6 | M_6)$ is the density of the uniform distribution over the interval $[-1, 1]$.

## 4. Empirical results

In this part we present an empirical example of Bayesian comparison of all competing specifications. As a basic dataset we considered $T = 1398$ observations of daily growth rates, $y_j$, of the WIBOR one month Zloty interest rate from 20.03.97 till 05.09.02. The variability of daily returns $y_j$ as well as some descriptive statistics are presented on Figure 1. It is clear, that dynamics of daily returns of the WIBOR1m instrument is very anomalous. Huge outliers, caused by changes in the monetary policy, together with the regions of almost no variability, depicts very volatile behavior of rates of daily changes of the Polish zloty middle term interest rate. In spite of the fact, that in five years from March 1997 to September 2002, the Polish money market was changing, our first attempt to compare all models was based on the whole dataset. As seen in Figure 1, negative value of the skewness statistics clearly shows substantial asymmetry of the empirical distribution. It also may indicate skewness of the conditional distribution of $y_j$.

In Table 1 we present the results of Bayesian comparison of explanatory power of all competing specifications. In rows we put the decimal logarithm of marginal data densities $p(y^{(t)} | M_i)$ ($i = 0, 1, \ldots, 6$), posterior probabilities of all models including $M_0$, posterior probabilities of all conditionally asymmetric GARCH specifications ($M_i$, $i = 1, \ldots, 6$) and Bayes factors of $M_0$ (representing conditional symmetry) against $M_i$, $i = 1, \ldots, 6$ (representing alternative i.e. conditional asymmetry). Both sets of $P(M_i|y^{(t)})$, for $i = 0, 1, \ldots, 6$ and $i = 1, \ldots, 6$ were obtained by imposing equal prior model probabilities.

It is clear, that the modeled dataset of daily returns of WIBOR1m interest rates do not support decisively superiority of any of competing skewing mechanism. The mass of posterior probabilities is rather dispersed among models. However, the greatest value if $P(M_i|y^{(t)})$ receives conditionally skewed Student-t GARCH model built on the basis of the hidden truncation idea. In this case the value of posterior probability is greater than 44%. The considered dataset also support conditionally skewed Student-t GARCH model with Beta distribution transformation ($M_3$) and conditionally $\alpha$-Stable GARCH specification ($M_6$). Those three models cumulate about 85% of the posterior probability mass, making all remained specifications (including conditionally symmetric $M_0$) rather improbable in the view of the data. Very small value of posterior probability is received by model $M_4$, which, just like $M_3$, is built on the basis of the Beta distribution transformation, but with two free parameters governing the type of skewness. The observed time series support parsimony of Beta distribution transformation with one free skewness parameter (in $M_3$) and rejects generalization proposed by [15]. Finally, model $M_4$ receives less than 9% of posterior probability mass. Also the Bernstein density transformation (with 2 free parameters) leads to very doubtful explanatory power of the resulting GARCH specification. The model $M_5$ is strongly rejected by the data, as the value of posterior probability is more than 10 times smaller than posterior probability of symmetric GARCH model ($M_0$). This may lead to the similar conclusion, as it was pointed by [11], that Bernstein densities do not yield flexible skewing mechanism for small values of m, see (6).

On the basis of posterior odds ratios $B_{0i}$ (for $i = 1, ..., 6$) we carried out Bayesian testing of conditional asymmetry within presented GARCH framework, according to the Jeffreys scale, see [13]. Except for $M_5$, posterior odds $P_{0i}$ reject conditional symmetry in favor of skewness of the conditional distribution of $y_j$ in model $M_i$. In case of model $M_2$ and $M_3$, the data strongly support conditional asymmetry, because $P_{02}$ and $P_{03}$ reaches the third grade of Jeffreys scale. The data substantially (grade 2) support $M_4$ and weakly (grade 1) support $M_1$ both against symmetric $M_0$. Additionally, poor explanatory power of specification $M_5$ is confirmed. The data strongly support (grade 3) symmetry against skewing mechanism built on the basis of Bernstein density transformation.

In Table 2 we present the results of Bayesian inference about tails and skewness of the conditional distribution of daily returns in all competing speci-fications. The tails of $p(y_j|\psi_{j-1},\theta,\omega_i,\eta_i,M_i)$ is modeled by the degrees of freedom parameter $\nu > 0$ in $M_i$, for $i = 0, 1, ..., 5$, while for $i = 6$ they are captured by the index of stability $\alpha \in (0, 2]$. Apart from making inference about model specific skewness parameters in all models, we also put posterior means and standard deviations of skewness measure $\gamma_M$, proposed by [2]. We also present values of posterior probability of left asymmetry of the density $p(y_j|\psi_{j-1},\theta,\omega_i,\eta_i,M_i)$ (i.e. $P(\gamma_M < 0|y^{(t)}, M_i)$).

In case of conditional symmetry (model $M_0$) the dataset clearly rejects the hypothesis of existence of the variance of the distribution $p(y_j|\psi_{j-1},\theta,\nu,M_0)$, because the whole density of the posterior distribution of the degrees of freedom parameter $\nu$ is located on the left side of the value $\nu = 2$. Also, very tight location of $p(\nu|y^{(t)}, M_0)$ around the value $\nu = 1.55$ in model $M_0$, assures that the conditional distribution of daily returns possesses first moment. Those properties of the posterior distribution $p(\nu|y^{(t)}, M_0)$ remains practically unchanged in case of the majority of conditionally skewed specifications. Only in model $M_4$, Beta distribution transformation with 2 free parameters substantially changes both, location and scale of the posterior density of the degrees of freedom parameter. In spite of the fact, that $p(\nu|y^{(t)}, M_4)$ is located on the right side of the value $\nu = 2$, the posterior standard deviation (equal to 0.55) leaves great uncertainty about existence of the second moment of the conditional distribution $p(y_j|\psi_{j-1}\theta,\nu,\eta_4,M_4)$. The similar conclusions can be drawn given model $M_6$, i.e. given the conditionally $\alpha$-Stable GARCH specification. Since the posterior mean of the index of stability $\alpha$ locates $p(\alpha|y^{(t)}, M_6)$ around the value $\alpha = 1.21$ (with posterior standard deviation 0.04), the dataset decisively rejects conditional normality in model $M_6$ (corresponding to $\alpha = 2$). From the definition of the family of the $\alpha$-Stable distributions the resulting conditional distribution $p(y_j|\psi_{j-1},\theta,\alpha,\eta_6,M_6)$ does not have variance (just like in $M_i$, $i = 0, 1, ..., 4$). Also, posterior distribution of $\alpha$ is located on the right side of the value $\alpha = 1$. It clearly assures the existence of conditional mean of the distribution of modeled daily returns (again just like in $M_i$, $i = 0, 1, ..., 5$). The posterior means and standard deviations of both, asymmetry parameters $\eta_i$ and skewness measure $\gamma_M$ indicate, that in all specifications $M_i$, $i = 1, \ldots, 6$ there is quite strong evidence in favor of left skewness of the conditional distribution of modeled daily returns. The posterior distributions of $\gamma_M$ are located on the left side of the value $\gamma_M = 0$, confirming left asymmetry of $p(y_j|\psi_{j-1},\theta,\omega_i,\eta_i,M_i)$. However, relatively great values of posterior standard deviations of $\gamma_M$ reduces potential strength of conditional skewness effect. As measured by posterior mean of $p(\gamma_M|y^{(t)}, M_i)$, the greatest intensification of skewness of $p(y_j|\psi_{j-1},\theta,\omega_i,\eta_i,M_i)$ is obtained in model $M_1$. In this case of GARCH model the posterior expectation of asymmetry measure is equal to $M = -0.063$, with posterior standard deviation equal about 0.033.

All remained conditionally skewed GARCH specifications generated posterior distributions of $\gamma_M$, localized much closer to the value $\gamma_M = 0$ and also much more dispersed. As a consequence, model $M_1$ yields the greatest value of posterior probability of left asymmetry of $p(y_j|\psi_{j-1}, \theta, \omega_i, \eta_i, M_i)$. In case of models $M_1, M_2, M_3, M_6$ the posterior probabilities $P(\gamma_M < 0|y^{(t)}, M_i)$ are greater than 91%. The conditionally skewed GARCH specification based on the Berstein density transformation ($M_5$) generates relatively low value of posterior probability of left asymmetry, making symmetry, as well as skewness to the right not strongly rejected by the data. Given Beta distribution transformation with two free parameters, posterior probability of $\gamma_M < 0$ is much lower than in case of model $M_3$. Again we may conclude, that generalization, based on two free parameters in Beta distribution, substantially changes inference about the properties of $p(y_j|\psi_{j-1}\theta, \omega_i, \eta_i, M_i)$. Finally, on the basis of the Bayesian model pooling technique, we obtained posterior probability of left asymmetry calculated considering the whole class of specifications $M_i$, $i = 1, ..., 6$. The modeled dataset clearly supports left asymmetry, as $P(\gamma_M < 0|y^{(t)}) = 0.9263$, but it also leaves some uncertainty about true intensification of this phenomenon. Posterior probability of symmetry and right skewness of $p(y_j|\psi_{j-1}, \theta, \omega_i, \eta_i, M_i)$ (equal to 0.0737) does not totally reject those cases.

## 5. Concluding remarks

The main goal of this paper was an application of Bayesian model comparison in testing the explanatory power of the set of competing GARCH (ang. Generalized Autoregressive Conditionally Heteroscedastic) specifications, all with asymmetric and heavy tailed conditional distributions. As an initial specification we considered GARCH process with conditional Student-$t$ distribution with unknown degrees of freedom parameter, proposed by [7]. By introducing skewness into Student-$t$ family and by incorporating the resulting class as a conditional distribution we generated various GARCH models, which compete in explaining possible asymmetry of both conditional and unconditional distribution of financial returns. In order to make Student-$t$ family skewed we considered various alternative methods recently proposed in the literature. In particular, we applied the hidden truncation mechanism (see [3], [1]), an approach based on the inverse scale factors in the positive and the negative orthant (see [10]), order statistics concept [14], two different settings of the Beta distribution transformation [15] and Bernstein density transformation (see [28]). Additionally, we presented the results of Bayesian inference within GARCH process with conditional $\alpha$-Stable distribution, (see [30], [29]).
Analysis of posterior probabilities of competing specifications did not lead to decisive conclusion about superiority of any of the considered specifications. The greatest value of $P(M_i|y^{(t)})$ received conditionally skewed Student-$t$ conditional distribution built on the basis of the hidden truncation mechanism (see [3]). The data also supported Beta distribution transformation with single free parameter and conditionally $\alpha$-Stable GARCH process. Those three models cumulated more than 85% of the posterior probability mass.

The results of Bayesian estimation showed, that in each competing specification the modeled data set confirmed left asymmetry of the conditional distribution $p(y_j|\psi_{j-1}, \theta, \omega_i, \eta_i, M_i)$. In all models $M_i$ ($i = 1, \ldots, 6$) the posterior distribution of skewness measure $\gamma_M$ was situated on the left side of the value $\gamma_M = 0$ (representing symmetry). However, substantial dispersion of $p(M|y^{(t)}, M_i)$, as measured by the posterior standard deviation of $\gamma_M$, did not preclude symmetry or right skewness of $p(y_j|\psi_{j-1}, \theta, \omega_i, \eta_i, M_i)$. As a result, the posterior probability of left asymmetry (equal to 0.9263), obtained by application of Bayesian model pooling approach, left some uncertainty about the true strength of conditional skewness phenomenon.

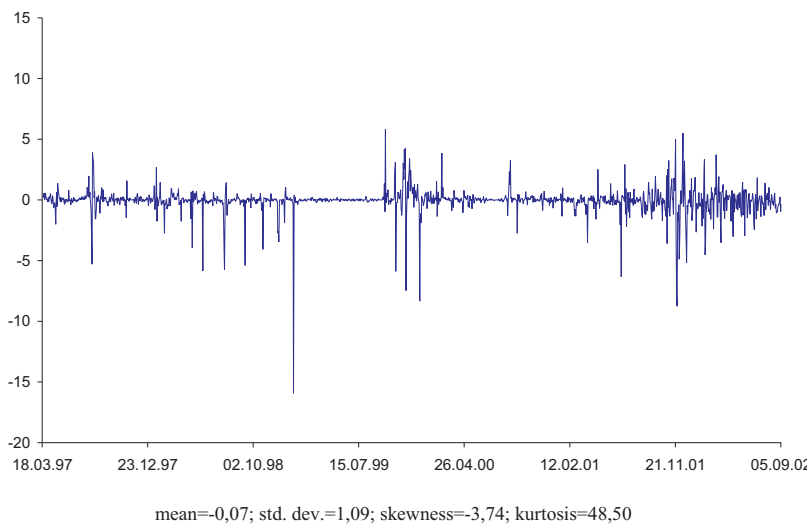

mean=-0,07; std. dev.=1,09; skewness=-3,74; kurtosis=48,50

Figure 1. Daily returns $y_j$ of the WIBOR 1-month Zloty interest rate from 20.03.1997 to 05.09.2002, $T = 1398$ observations

|  | $M_2$ Azzalini (1985), $\gamma_2 \in \mathbb{R}$ $\gamma_2 \sim N(0,3)$ | $M_3$ Beta distribution with one parameter, Jones (2004), $\gamma_3 \in (0,+\infty)$ $\ln\gamma_3 \sim N(0,1)$ | $M_4$ Beta distribution with two parameters, Jones and Faddy (2003), $a \in (0,+\infty)$, $b \in (0,+\infty)$ $\ln a \sim N(0,1)$, $\ln b \sim N(0,1)$ | $M_5$ Bernstein densities, 2 parameters, $w_1 \sim N(0.33;36)$, $w_2 \sim N(0.33;36)$, $w_1>0$, $w_2>0$, $w_1+w_2<1$ | $M_1$ Fernández and Steel (1998) $\gamma_1 \in (0.5; 2)$ $\ln\gamma_1 \sim N(0,1)$ | $M_6$ ($\alpha$-Stable GARCH) $\beta \in (-1,1)$ $\beta \sim U(-1,1)$ | $M_0$ Symmetric Student-$t$ GARCH |
|---|---|---|---|---|---|---|---|
| $\log p(y^{(t)}|M_i)$ | -356.40 | -356.62 | -357.11 | -358.78 | -357.42 | -356.87 | -357.69 |
| $P(M_i|y^{(t)})$, $i=0,...,6$ | 0.437 | 0.262 | 0.086 | 0.002 | 0.041 | 0.150 | 0.023 |
| $P(M_i|y^{(t)})$, $i=1,...,6$ | 0.447 | 0.268 | 0.088 | 0.002 | 0.042 | 0.153 | x |
| $P_{0i}$ | 0.052 | 0.086 | 0.263 | 12.5 | 0.549 | 0.151 | 1 |
| Jeffreys grade | Strong (3) | Strong (3) | Substantial (2) | Strong (3) against $M_5$ in favour o $M_0$ | Weak (1) | Substantial (2) | x |

Table 1. Logarithms of marginal data densities, posterior probabilities of all competing models (including $M_0$) and Bayes factors of $M_0$ against $M_i$, for $i =$

*Bayesian.tex    printed on 21st August 2018*    17|  | $M_2$ Azzalini (1985), $\gamma_2 \in R$ $\gamma_2 \sim N(0,3)$ | $M_3$ Beta distribution with one parameter, Jones (2004), $\gamma_3 \in (0,+\infty)$ $\ln\gamma_3 \sim N(0,1)$ | $M_4$ Beta distribution with two parameters, Jones and Faddy (2003), $a \in (0,+\infty)$, $b \in (0,+\infty)$ $\ln a \sim N(0,1)$, $\ln b \sim N(0,1)$ | $M_5$ Bernstein densities, 2 parameters, $w_1 \sim N(0.33;36)$, $w_2 \sim N(0.33;36)$, $w_1 > 0$, $w_2 > 0$, $w_1 + w_2 < 1$ | $M_1$ Fernández and Steel (1998) $\gamma_1 \in (0.5; 2)$ $\ln\gamma_1 \sim N(0,1)$ | $M_6$ ($\alpha$-Stable GARCH) $\beta \in (-1,1)$ $\beta \sim U(-1,1)$ | $M_0$ Symmetric Student-$t$ GARCH |
|---|---|---|---|---|---|---|---|
| Symmetry | $\gamma_2 = 0$ | $\gamma_3 = 1$ | $a = b = 1$ | $w_1 = w_2 = 1/3$ | $\gamma_1 = 1$ | $\beta = 0$ | always |
| tail parameters | $\nu$  1.55  *0.10* | $\nu$  1.59  *0.10* | $\nu$  2.07  *0.55* | $\nu$  1.54  *0.20* | $\nu$  1.58  *0.10* | $\alpha$  1.21  *0.04* | $\nu$  1.55  *0.10* |
| asymmetry parameters $\eta_i$ | $\gamma_2$: -0.046  *0.018* | $\gamma_3$: 0.942  *0.020* | $a$: 0.951  *0.100* $b$: 1.070  *0.091* | $w_1$: 0.493  *0.191* $w_2$: 0.255  *0.276* | $\gamma_1$: 0.939  *0.031* | $\beta$: -0.017  *0.011* | x |
| $\gamma_M$, symmetry $\gamma_M = 0$ | -0.027  *0.018* | -0.041  *0.030* | -0.035  *0.040* | -0.060  *0.075* | -0.063  *0.033* | -0.015  *0.010* | x |
| $P(\gamma_M < 0 \| y^{(t)}, M_i)$ | 0.9348 | 0.9184 | 0.8610 | 0.7872 | 0.9756 | 0.9446 | x |
| $P(\gamma_M < 0 \| y^{(t)})$ | | | | 0.9263 | | | x |

Table 2. Posterior means and standard deviations of tails and asymmetry parameters in all models as well as posterior probability of left skewness of